\documentclass[12pt]{iopart}
\usepackage{graphicx}
\usepackage{subfig}
\usepackage{floatrow}
\usepackage{hyperref}
\begin{document}

\title[Absolute efficiency of a two-stage microchannel plate for electrons]{Absolute efficiency of a two-stage microchannel plate for electrons in the 30 - 900 eV energy range}

\author{A Apponi$^1$\footnote{Corresponding author.}, F Pandolfi$^2$, I Rago$^2$, G Cavoto$^3$, C Mariani$^3$ and A Ruocco$^1$}

\address{$^1$ Universit\`a di Roma Tre e INFN Sezione di Roma Tre, Via della Vasca Navale 84, 00146 Rome, Italy}
\address{$^2$ INFN Sezione di Roma, Piazzale Aldo Moro 2, 00185 Rome, Italy}
\address{$^3$ Sapienza Universit\`a di Roma e INFN Sezione di Roma, Piazzale Aldo Moro 2, 00185 Rome, Italy}

\ead{alice.apponi@uniroma3.it}

\begin{abstract}
We report on an apparatus able to measure the absolute detection efficiency of a detector for electrons in the 30 - 900 eV range. In particular, we discuss the characterisation of a two-stage chevron microchannel plate (MCP). The measurements have been performed in the LASEC laboratory at Roma Tre University, whit a custom-made electron gun. The very good stability of the beam current in the fA range, together with the picoammeter nominal resolution of 0.01 fA, allowed the measurement of the MCP absolute efficiency $\epsilon$. We found an $\epsilon = (0.489 \pm 0.003)$ with no evident energy dependence. We fully characterised the MCP pulse shape distribution, which is quasi-Gaussian with a well visible peak above the noise level. We measured a 68$\%$ variation of the average pulse height between 30 and 500 eV. Furthermore, with a deeper analysis of the pulse shape, and in particular of the correlation between pulse height, area and width, we found a method to discriminate single- and multi- electron events occurring within a 10 ns time window.
\end{abstract}


This is the version of the article before peer review or editing, as submitted by an author to Measurement Science and Technology. IOP Publishing Ltd is not responsible for any errors or omissions in this version of the manuscript or any version derived from it. The Version of Record is available online at \url{https://doi.org/10.1088/1361-6501/ac3d07}.

\maketitle

\section{Introduction}
The characterisation of an electron detector, and in particular its absolute efficiency, is crucial in the context of the development of new experiments which rely on electrons as a final product. Some examples are a tritium-based telescope for the detection of the cosmological neutrino background (PTOLEMY \cite{PTOLEMY3, PTOLEMY4}) and a revisitation of a photomultiplier, in which the cathode is an array of vertically-aligned carbon nanotubes, for the detection of light dark matter and UV light (Dark-PMT and NanoUV projects \cite{Cavoto, Apponi:2021lyd, Cavoto:2019flp}). The aim of the PTOLEMY project is to study the endpoint of the $\beta$-electron spectrum, emitted by the tritium decay \cite{Cocco}. On the other hand, a cathode made of vertically-aligned carbon nanotubes is an anisotropic target which is sensitive to UV light and, eventually, light dark matter (sub-GeV). The anisotropy of the target is the key aspect to account for the directionality in the detection of the light dark matter and an efficient counting detector for the electrons is crucial as well.\\
For electron detectors in single electron counting mode, the absolute efficiency can be measured as the ratio of the output count rate, times the electron charge, and the input current emitted by the electron source. Therefore, to measure the detector absolute efficiency, the basic requirement is to have a range of intersection between the operating limits of the detector, the source and the ammeter, which measures the source current. Typically, the maximum count rate of an electron detector is of the order of 1 MHz, then the electron source must be able to provide currents below $\sim$100 fA and the picoammeter must have the resolution to measure them.\\
With our apparatus it is possible to perform measurements of absolute efficiency by using as electron source a custom-made monochromatic electron gun. This electron gun provides stable beam currents in the range 5 fA - 100 nA, with energy tuneable between 30 and 900 eV. The current is measured through a Faraday cup connected to a picoammeter, which has a minimum resolution of 0.01 fA.\\
The same apparatus was previously employed to characterise the response of a windowless silicon APD (avalanche photo-diode) to low-energy electrons \cite{APD}. In that case, we did not measure the single electron counts at the output, as in our set-up the noise level was too high, but the current generated by the APD. We observed that the output current was proportional to the input current. The proportionality factor is the APD gain, which includes the signal amplification and the efficiency.\\
In this paper, we report on the characterisation of a microchannel plate (MCP) which is operated in single electron counting mode. The MCPs are widely used for the detection of charged particles and photons. Originally developed as electron multipliers for image intensification, MCPs are employed in time-of-flight spectrometers \cite{Tulej, Kennerly, Dhawan} or imaging X-ray astronomy \cite{FraserXray}, as well as in electron analysers for photoemission spectroscopy and many other applications. The measurements of the MCPs absolute efficiency for electrons down to the low-energy region reported so far are still few, Galanti et al. \cite{Galanti} (0.05 - 10 keV) and M\"{u}ller et al. \cite{Muller} (0.1 - 2.3 keV), while other works either focus at higher energies \cite{Tulej} or contain relative efficiency measurements \cite{Kennerly, Goruganthu}.\\
We therefore present here the measurement of the absolute efficiency of a two-stage chevron MCP for electrons in the 30 - 900 eV energy range, and the full characterisation of the pulse height distribution as a function of the input energy and current. Moreover, with a further study of the signal pulse shape, we also found a method to discriminate single- and multi- electron events which occur in a time window of 10 ns.

\section{Experimental set-up}
\begin{figure}[h]
\centering
\sidesubfloat[]{\label{fig:MCPontherod} \includegraphics[scale=0.2]{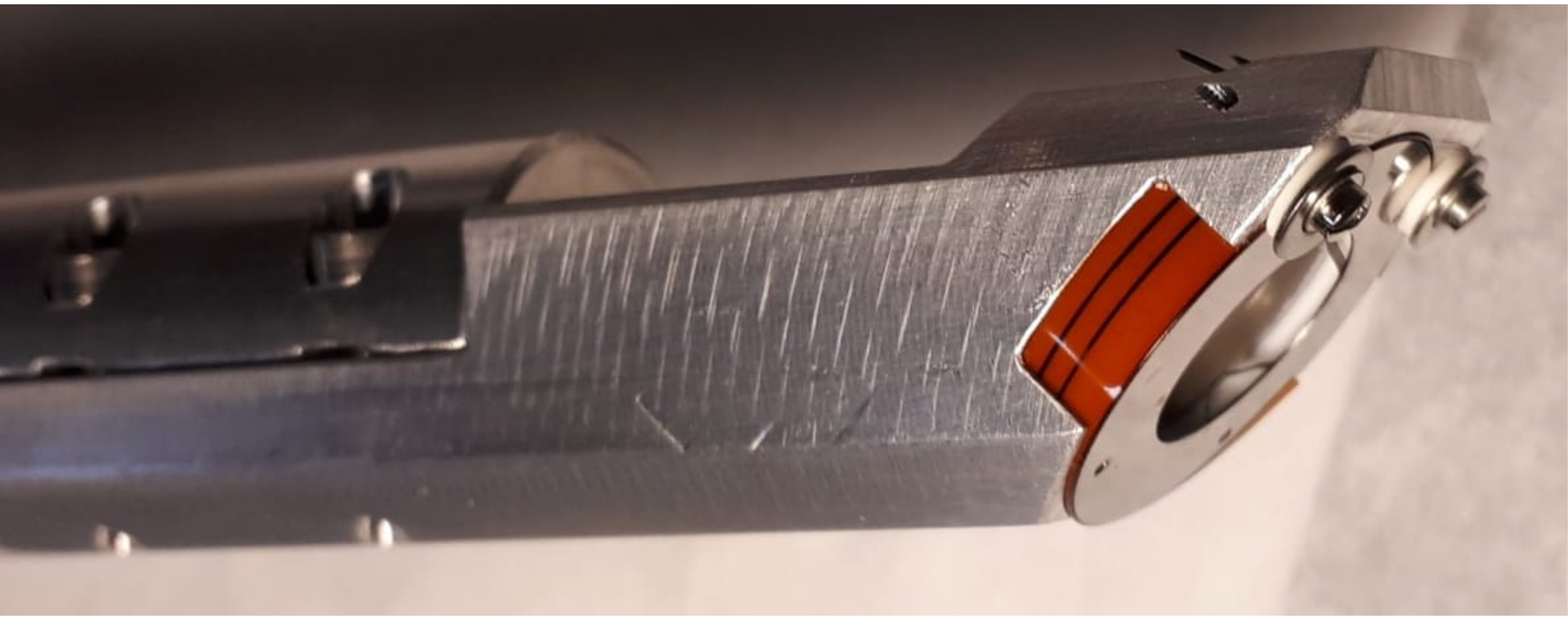} }
\quad
\sidesubfloat[]{\label{fig:circuit} \includegraphics[scale=0.4]{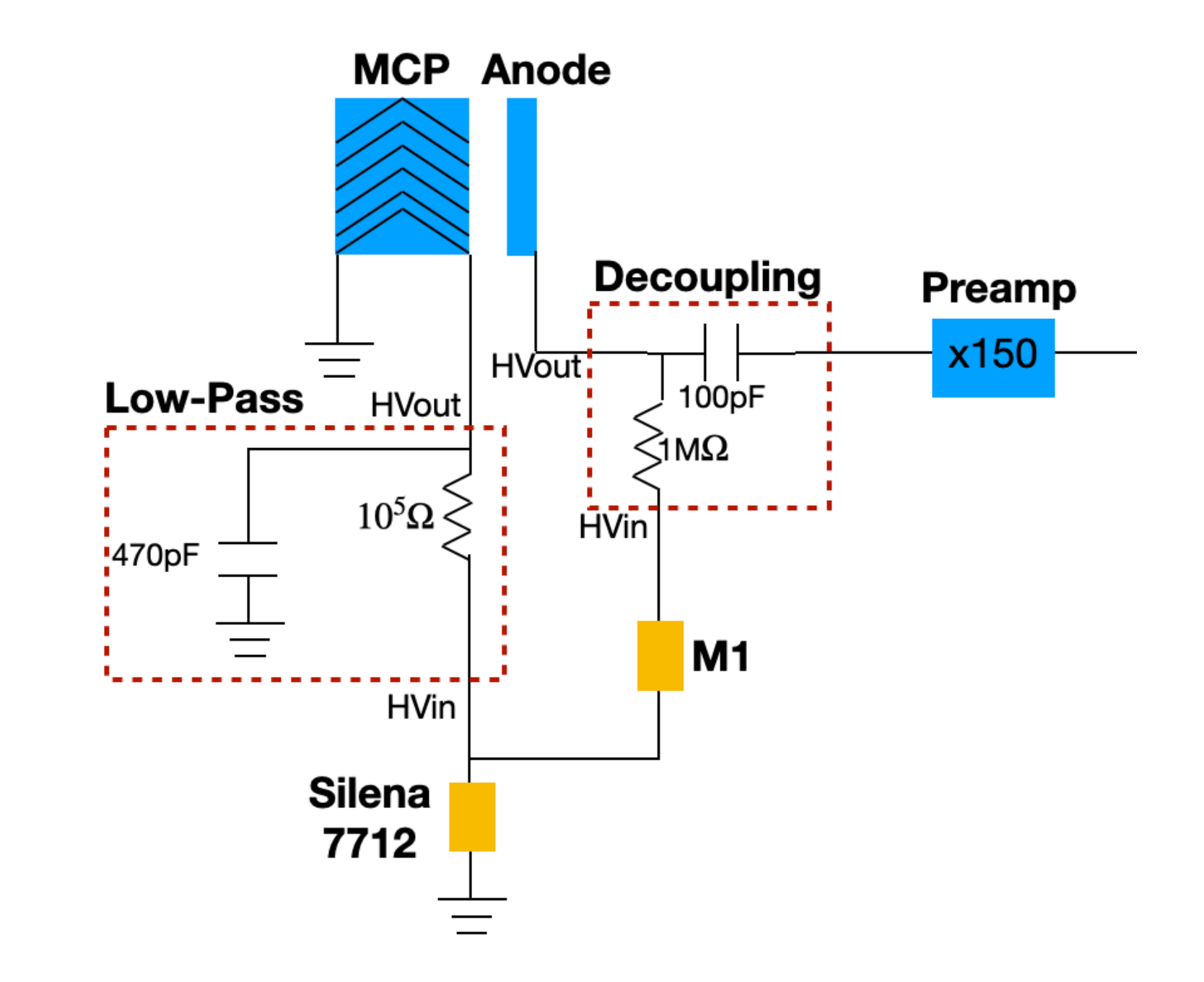} }
\caption{(a) Picture of the support rod with the MCP mounted on. (b) A schematic diagram of the MCP polarisation and read-out circuit. In the scheme, "M1" is the Mesar floating power supply.} 
\label{fig:MCPcircuitandrod}
\end{figure}

The apparatus for the absolute efficiency measurements is hosted in the ultra-high vacuum (UHV) chamber of the LASEC laboratory at Roma Tre University. The base pressure of the experimental chamber is in the low $10^{-10}$ mbar range. The electron gun is inserted in the UHV chamber at a fixed position. The detector to be tested is mounted on a custom-made support rod (Figure \ref{fig:MCPontherod}), which can translate inside the chamber through a linear shifter, and the Faraday cup is mounted on a motorised manipulator. This set-up allows either to bring the detector in the electron gun line of sight or to pull it back and insert the Faraday cup to perform a precise current measurement.

\begin{figure}[h]
\centering
\sidesubfloat[]{\label{fig:FCscan} \includegraphics[scale=0.28]{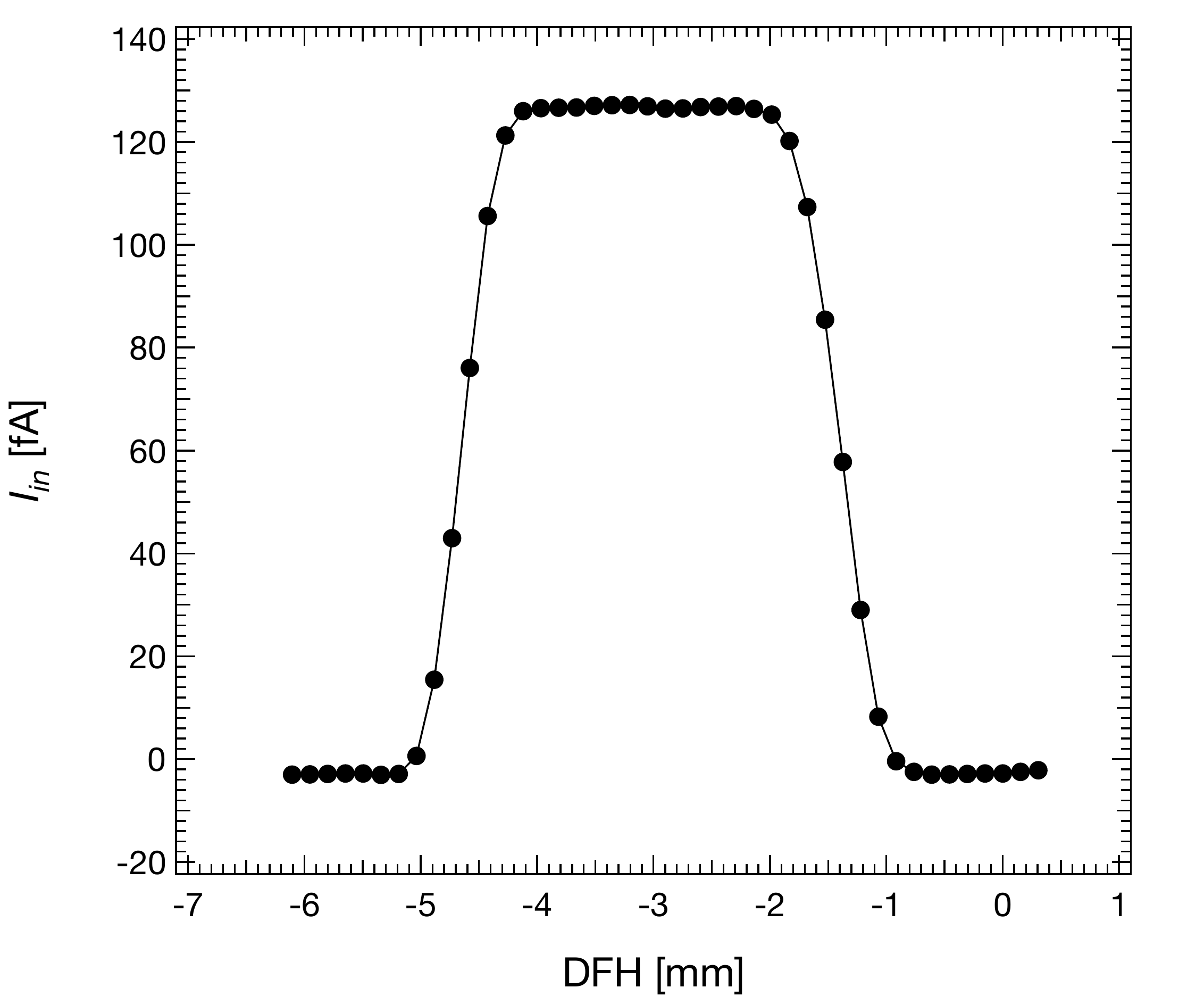} }
\quad
\sidesubfloat[]{\label{fig:FCscanderivative} \includegraphics[scale=0.28]{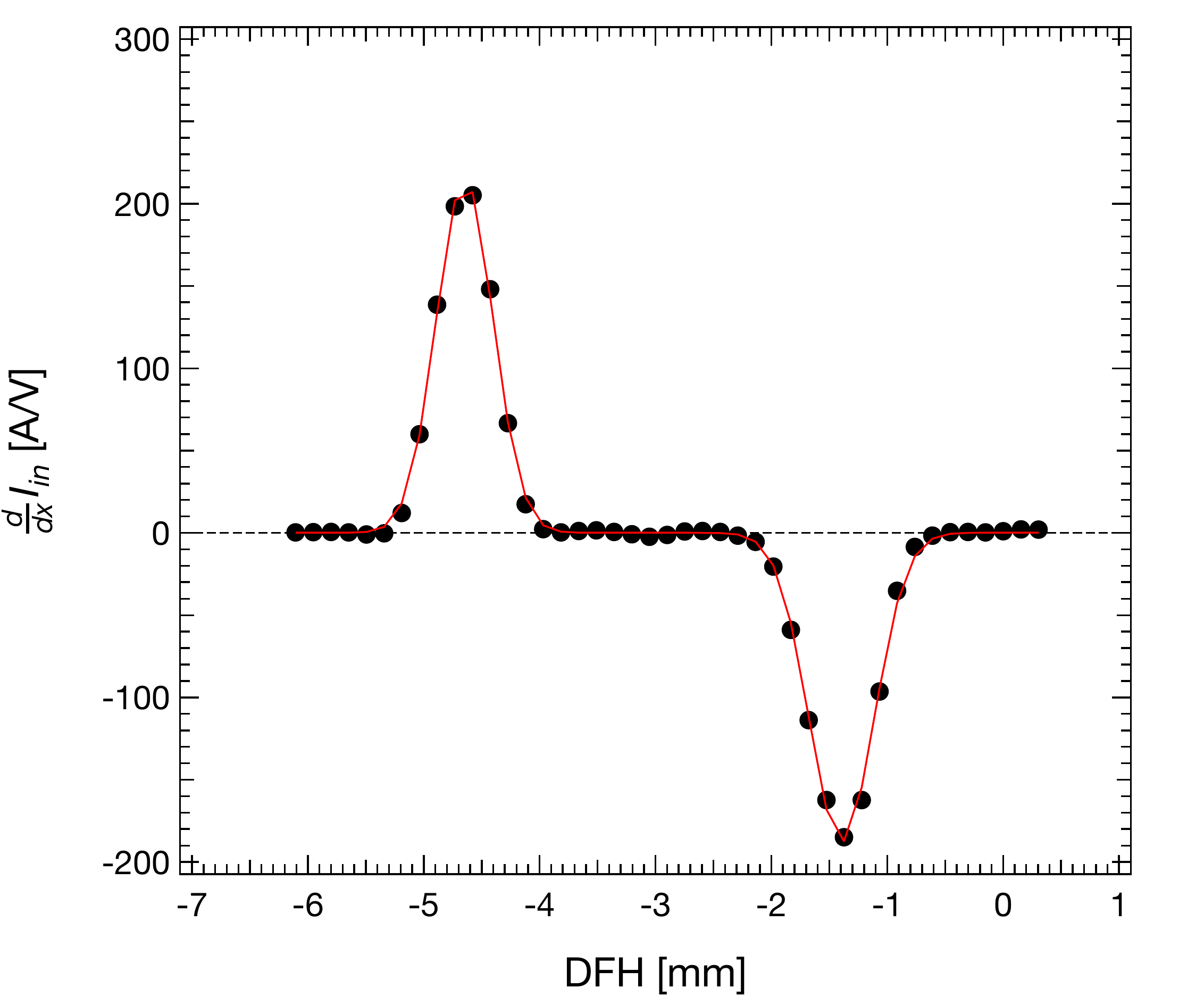} }
\caption{(a) Current profile measured during the scan of the electron beam on the Faraday cup hole. (b) Derivative of the current profile.} 
\label{fig:currentprofile}
\end{figure}

The electron gun is a monochromatic custom-made source of electrons operating in continuous current mode \cite{DiFilippo_Liscio_Ruocco_2020, Paoloni_DiFilippo_Cvetko_Kladnik_Morgante_Ruocco_2020}. It consists of three basic units: the source, the energy selection and the acceleration zone. The electrons are produced by a hot tungsten filament via thermionic emission with an energy spread of $\sim$400 meV. The beam is then focused toward two concentric hemispherical electrodes which reduce the energy dispersion down to 45 meV. Finally, through a series of properly polarised electrodes, the electron beam can be accelerated in the 30 - 900 eV range.\\ 
The electrodes in the acceleration region are also responsible for the beam focusing and its two-dimensional deflection. The size of the beam can be varied and then measured \emph{in situ} by exploiting the entrance hole of the Faraday cup. The beam is progressively deflected from side to side of the hole, which has a diameter of 3 mm. The current measured during this scan results to be the convolution of two profiles: the Gaussian of the beam and the step-like of the 3 mm wide hole. In Figure \ref{fig:FCscan}, a current profile measured for a beam of energy 264 eV and current 127 fA is shown. The size of the beam is evaluated from the derivative of the current profile, shown in Figure \ref{fig:FCscanderivative}, by fitting the curve with two Gaussian distributions. The full width at half maximum (FWHM) represents the beam size and in the shown case the result is FWHM = ($0.61 \pm 0.01$) mm.

\begin{figure}[h]
\centering
\sidesubfloat[]{\label{fig:focused} \includegraphics[scale=0.24]{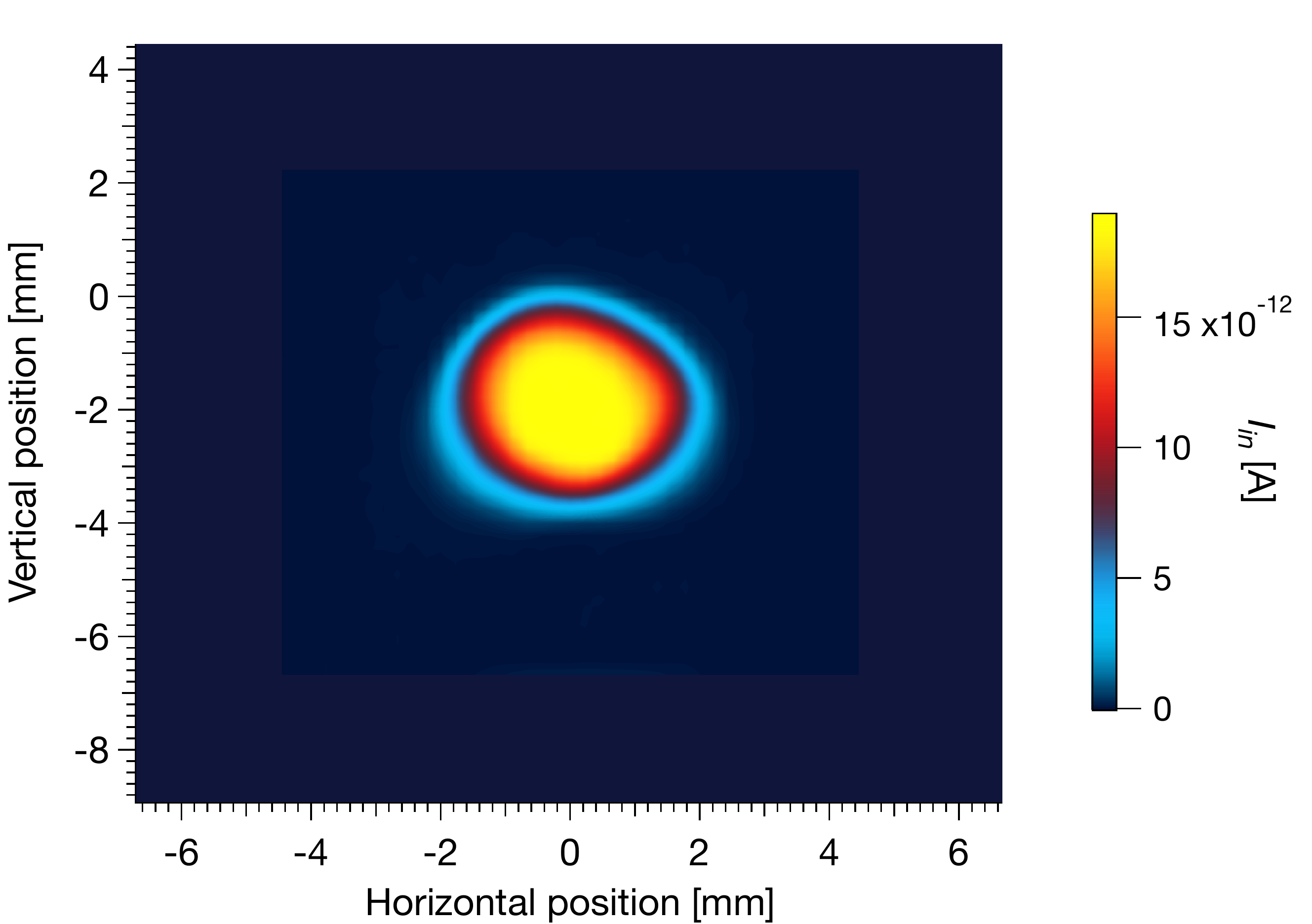} }
\quad
\sidesubfloat[]{\label{fig:large} \includegraphics[scale=0.24]{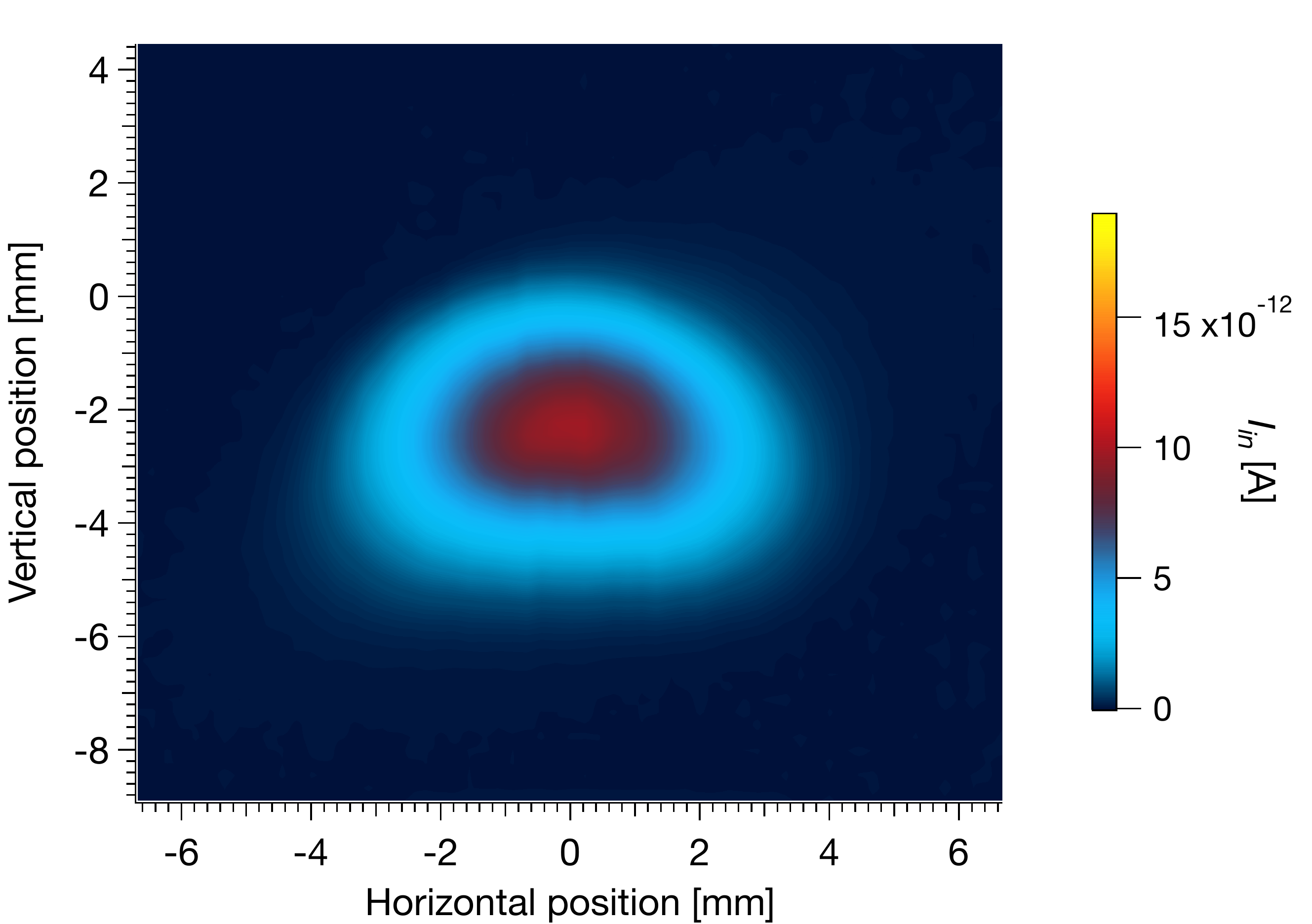} }
\caption{Electron current as a function of the beam position on the Faraday cup along two orthogonal directions. (a) Focused beam, (b) large beam.} 
\label{fig:totalcurrent}
\end{figure}

It is important to underline here that the variation of the beam size modifies the current density only, while the total electron current remains the same. In Figure \ref{fig:totalcurrent}, the 2D maps of the beam at energy 92 eV are shown. A focused beam (Figure \ref{fig:focused}) of diameter approximately 1 mm and a larger beam (Figure \ref{fig:large}) of diameter approximately 3 mm. The 2D maps have been obtained by repeating scans of the beam across the Faraday cup hole along one direction at positions gradually changed along the orthogonal direction. After the picoammeter zero subtraction, the sum of the current measured in each point of the 2D scan is 3.2 nA for focused beam and 3.3 nA for large beam. The total current is approximately equal for the two beam sizes, thus we can conclude that only the current density has changed.

The electron beam current can be tuned from 100 nA down to a few fA and is measured through the Faraday cup with a B2987A Keysight picoammeter (0.01 fA resolution in the 2 pA measurement range). \\
We characterised a commercial MCP by Hamamatsu Photonics (F1551-21S). It is a two-stage circular MCP in chevron configuration with channel bias angle of about $8^\circ$. The MCP outer diameter is 27 mm and the diameter of the effective area is 14.5 mm. The ratio of the channels open area to the effective area of the MCP is called open area ratio (OAR) and it is of 60$\%$ for the MCP we tested, with channels diameter of 12 $\mu$m. According to the final test sheet provided by Hamamatsu, this MCP has a 473.0 M$\Omega$ resistance and a minimum gain of $1\cdot10^6$, measured at supply voltage of 2 kV and ambient temperature of $25^\circ$ C.\\
For the tests reported in this paper, the MCP was polarised with a Silena 7712 power supply connected to the MCP back electrode, at nominal voltage of 1.8 kV, and with a Mesar floating power supply which provided a polarisation difference of +100 V to the anode. The MCP front was grounded in order to avoid electric fields in the fly region of the electrons, from the electron gun output, also grounded, to the MCP input. A schematic diagram of the polarisation and read-out circuit can be seen in Figure \ref{fig:circuit}. The output signals are negative pulses and, through the preamplifier (Ortec VT120), the pulse amplitude is multiplied by a factor 150. For the characterisation of the MCP, either the measurement of the output count rate or the pulse shape analysis were studied. Due to the occurrence of saturation effects in the MCP at high input currents \cite{Schagen}, the characterisation of the device has been limited from a few fA up to approximately 280 fA electron beam current. To measure the output count rate, the amplified signal was driven to a standard counting chain, where an edge discriminator selected the signal above a threshold set to -120 mV. On the other hand, the analysis of the pulse shape has been performed with a RTA4004 Rohde $\&$ Schwarz digital oscilloscope connected to the preamplifier output. The trigger level was set equal to the discriminator threshold used for the count rate measurements. The bandwidth used for the measurements presented in this paper is 1 GHz with 5 GSa/s sampling rate. We characterised the pulse height distribution as a function of the two beam parameters, current and energy, and for this purpose we exploited the Automatic Measurement tool of the oscilloscope. The minimum amplitude value within the displayed waveform was recorded. The uncertainty on this measurement, from Rohde $\&$ Schwarz specifications, is 1$\%$ + 50.5 mV. For a further analysis of the pulse shape, we also measured the pulse width and area, looking for a correlation with the pulse height. In the first case, the time between a falling edge and the following rising edge was measured at half maximum. For the pulse area, the integral of the displayed waveform was acquired.\\

\section{Pulse shape analysis}
\label{sec:pulseshape}
\begin{figure}[h]
\centering
\includegraphics[scale=0.3]{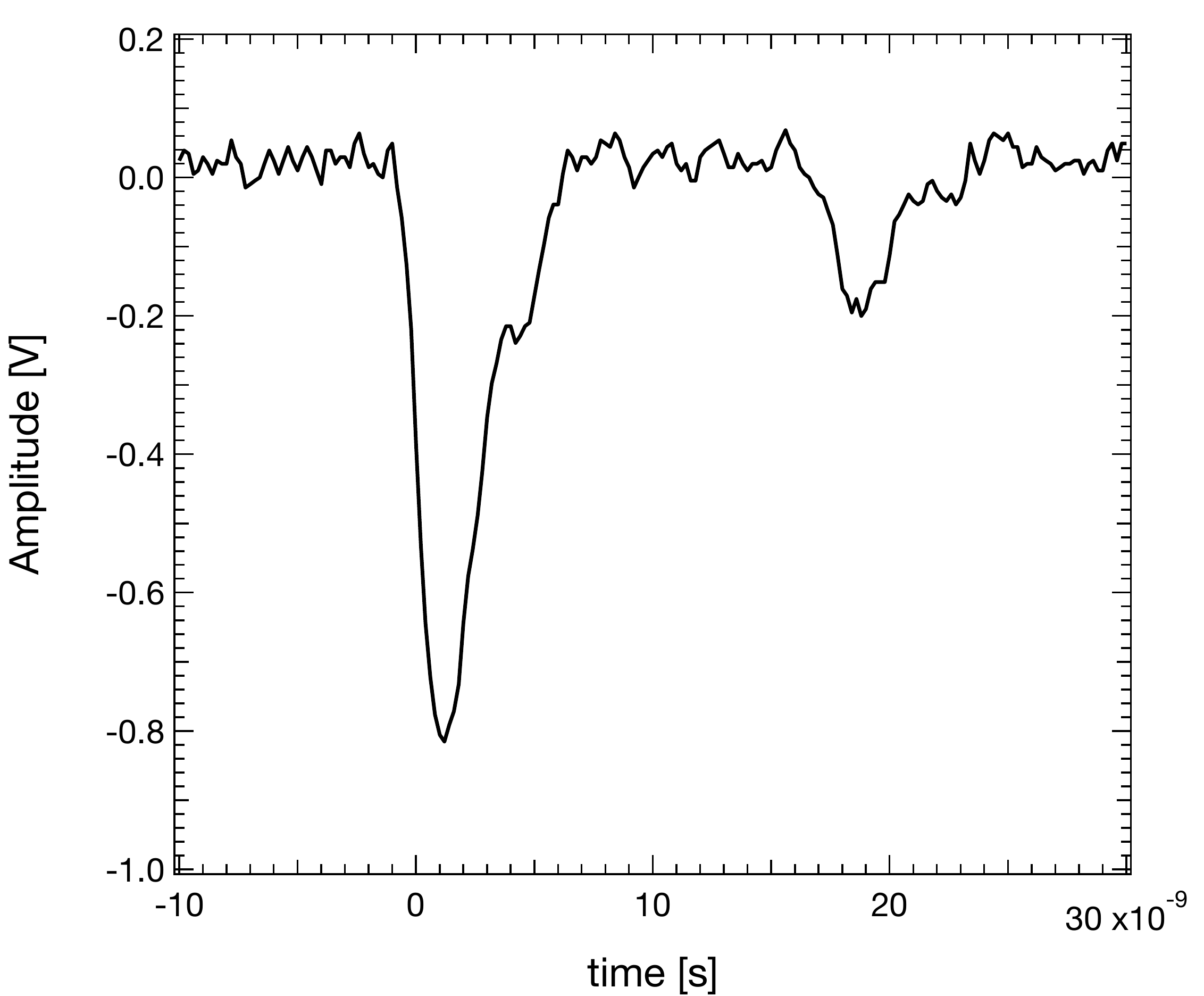}
\caption{A typical MCP waveform.}
\label{fig:waveform}
\end{figure}

A typical MCP output waveform is shown in Figure \ref{fig:waveform}. In the displayed waveform two peaks can be identified: a main pulse and a smaller one. The latter is a reflection always present at approximately 17 ns from the main pulse and due to the long cable in vacuum along the support rod of the MCP. In order to cut off the reflection, the time window of the oscilloscope chosen for the pulse shape analysis is $\tau$ = 10 ns after trigger, which correspond to the time position zero. The typical pulse had a fast rise time of 2 ns, a slower fall time of 5 ns and a width of approximately 3 ns (half maximum), while the pulse height distribution depends on the conditions of the electron beam sent on the MCP, such as the current and the energy.

\begin{figure}[h]
\centering
\includegraphics[scale=0.43]{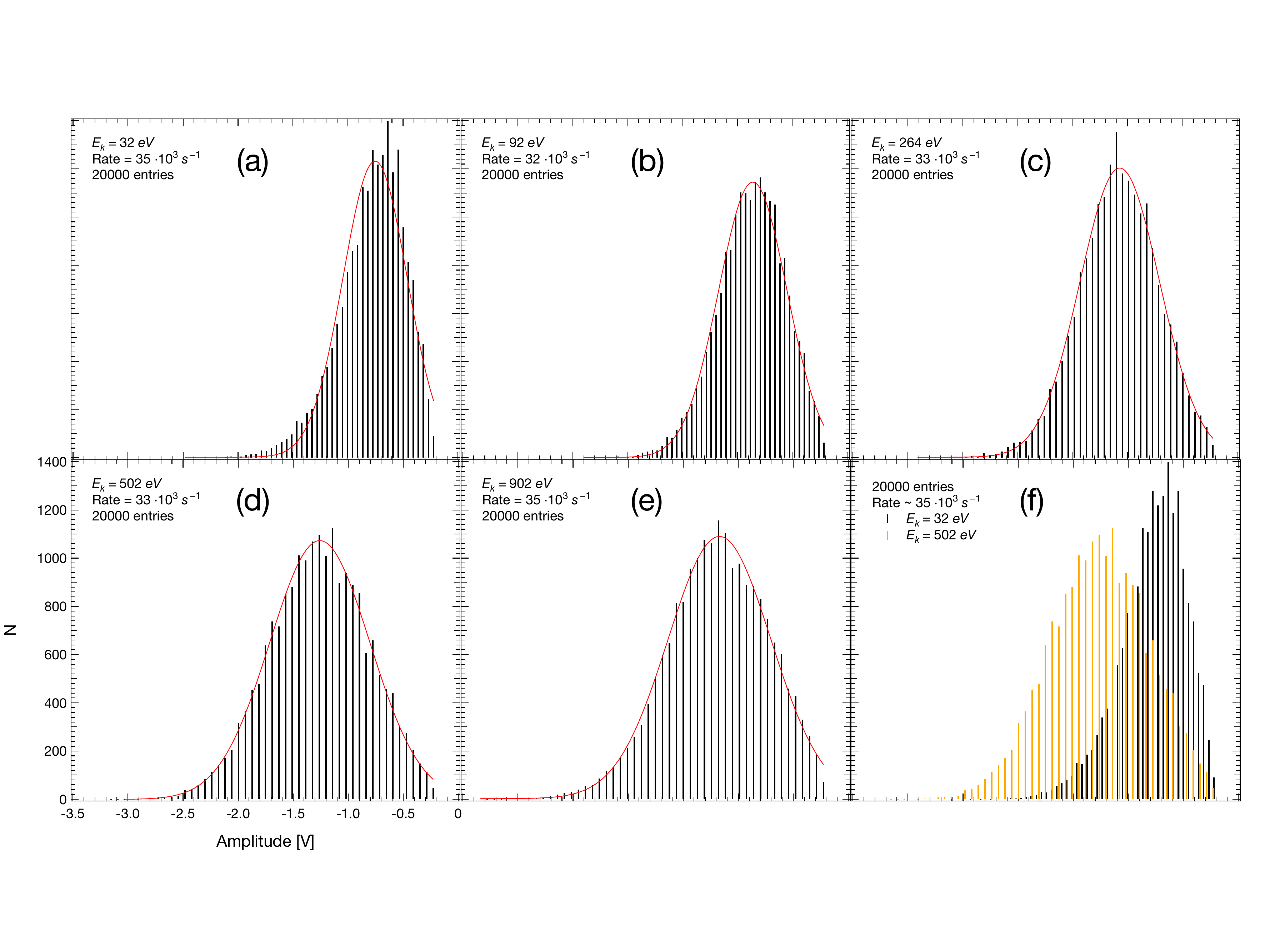}
\caption{Pulse height distributions at different electron energies $E_k$: (a) 32 eV, (b) 92 eV, (c) 264 eV, (d) 502 eV, (e) 902 eV. The red line is the result of the Gaussian fit. In (f) the PHDs at 32 and 502 eV are plotted together for comparison. For all measurements the trigger level is set to -120 mV, equal to the discriminator threshold in the counting chain. The x and y scales are the same for all the plots.}
\label{fig:PHDs}
\end{figure}

The typical pulse height distribution (PHD) of a chevron MCP has a quasi-Gaussian shape \cite{Muller, Goruganthu}, differently from a straight MCP in which the distribution is a negative exponential \cite{Galanti}. The difference is related to the suppression of the ion feedback within the channel, due to the chevron configuration, and the resulting increase of the gain. A dynamic equilibrium is reached between the space charge density in the last part of the channel and the kinetic energy of the electrons \cite{Wiza}. \\
We studied the pulse height distribution as a function of the electron energy in the 30-900 eV range, by acquiring the pulse amplitude on samples of 20000 waveforms with the oscilloscope. In Figure \ref{fig:PHDs}, the histograms of the pulse height are shown from (a) to (e) for beam energy 32, 92, 264, 502, 902 eV ($\pm 0.5$ eV) at approximately the same beam current of 10 fA. The PHDs at 32 and 502 eV are plotted together in (f) for comparison. The distributions are quasi-Gaussian, as expected, and the peak is well visibile, because sufficiently far from the discriminator threshold. From the plot in (f), a separation between the two distributions can be seen, which stands for a weak energy resolution of the device. In order to characterise the energy dependence of the pulse amplitude mean value, thus the energy resolution of the MCP, we fitted the histograms with Gaussian distributions (red lines). 

\begin{figure}[h]
\centering
\sidesubfloat[]{\label{fig:energyresolution} \includegraphics[scale=0.26]{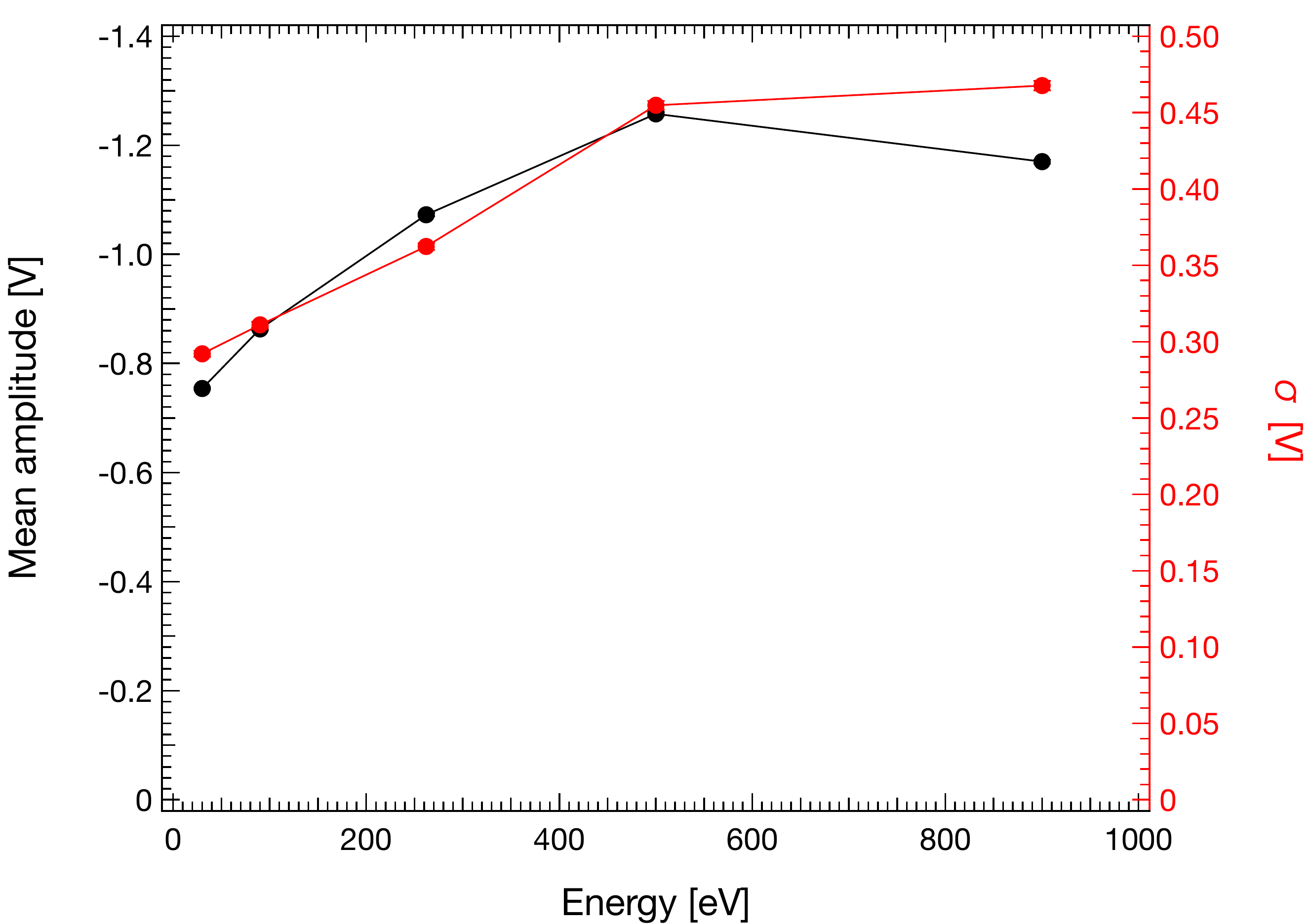} }
\quad
\sidesubfloat[]{\label{fig:PHR} \includegraphics[scale=0.26]{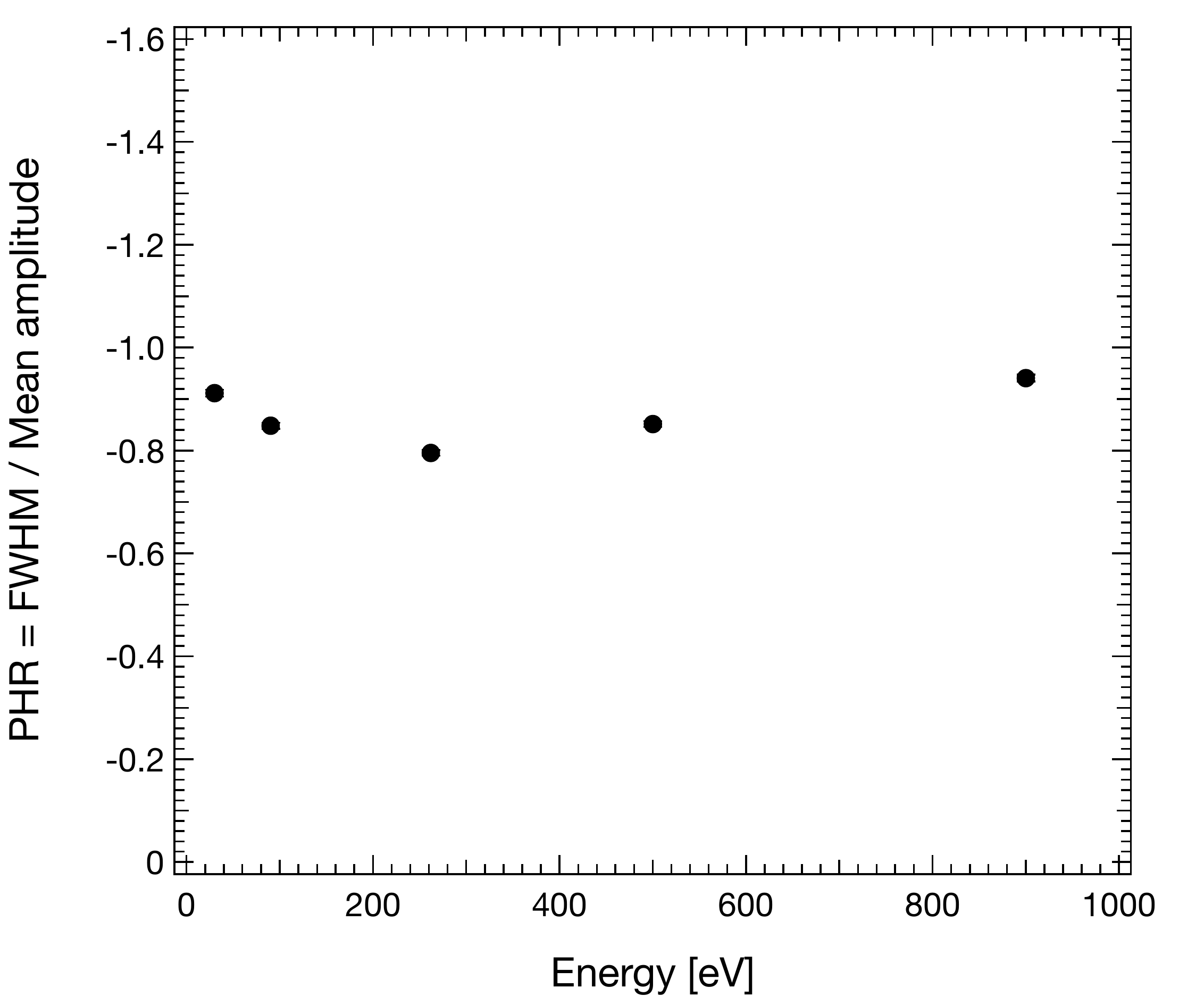} }
\caption{(a) Mean amplitude (in black) and $\sigma$ (in red) as a function of the electron energy, obtained from the Gaussian fit on the PHDs. (b) Pulse height resolution (PHR) as a function of the electron energy.} 
\label{fig:reslution}
\end{figure}

The fit result can be seen in Figure \ref{fig:energyresolution}, where the mean value (in black) and $\sigma$ (in red) of the distributions are plotted as a function of the beam energy. We found that the mean amplitude grows by 68$\%$ from 32 to 502 eV, where it reaches the maximum, and then slightly decreases at higher energy. The value of $\sigma$ increases within the energy range, with a slower growth between 502 and 902 eV.\\
The pulse height resolution (PHR) of the MCP, obtained from the full width at half maximum over the mean amplitude of the PHD (absolute value), has been computed and is reported in Figure \ref{fig:PHR} for all the energies in the experimental range. We found a PHR minimum value at 264 eV of 79$\%$.

\begin{figure}[h]
\centering
\includegraphics[scale=0.43]{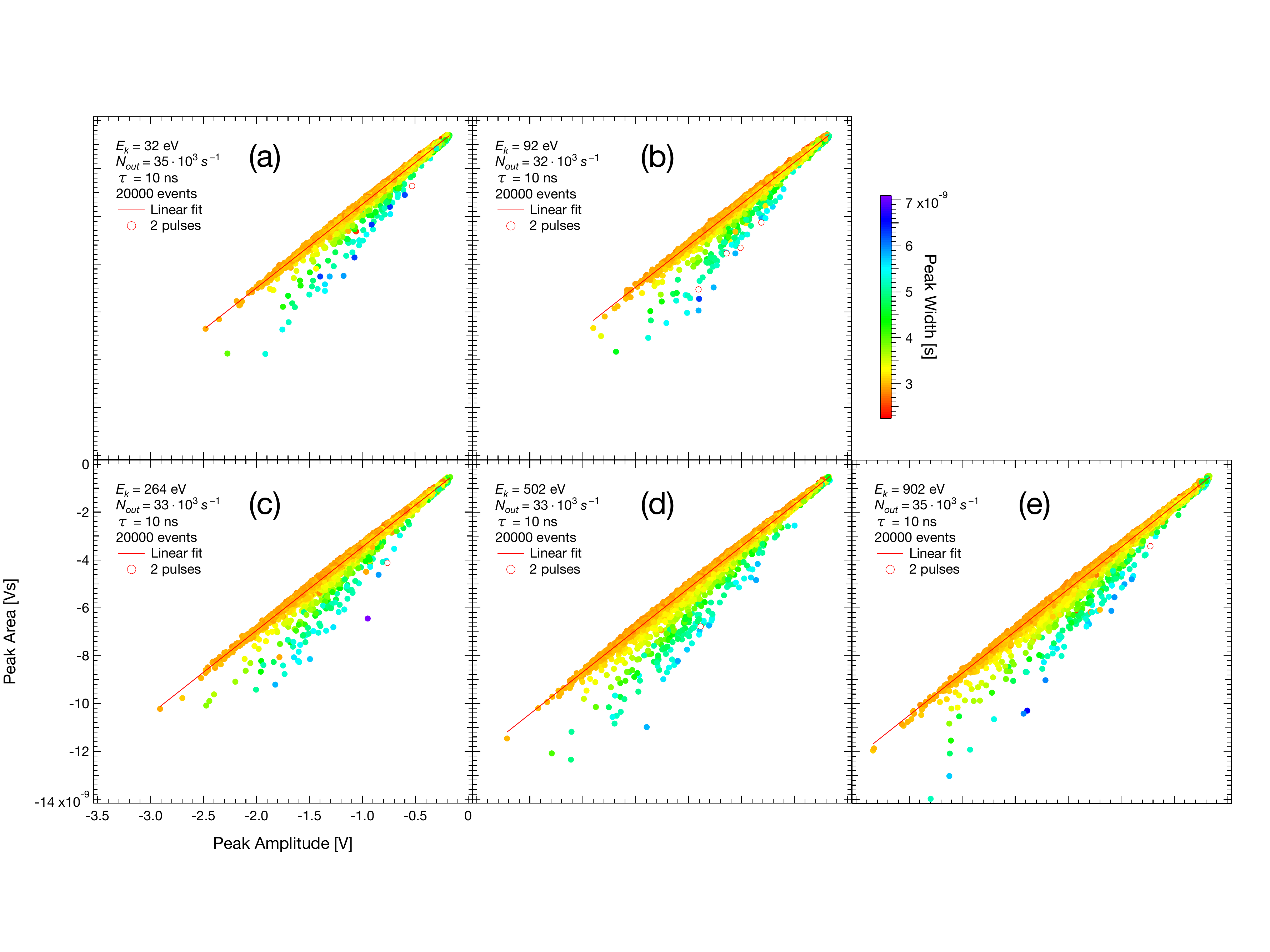}
\caption{Peak area as a function of the amplitude, the colour scale represents the peak width. The plots refer to different electron energies $E_k$: (a) 32 eV , (b) 92 eV , (c) 264 eV, (d) 502 eV, (e) 902 eV. The events in which two separated peaks has been identified are reported as red open markers. The solid red line is the result of the linear fit. The x, y and z scales are the same for all the plots.}
\label{fig:Correlations}
\end{figure}

\begin{figure}[h]
\centering
\includegraphics[scale=0.3]{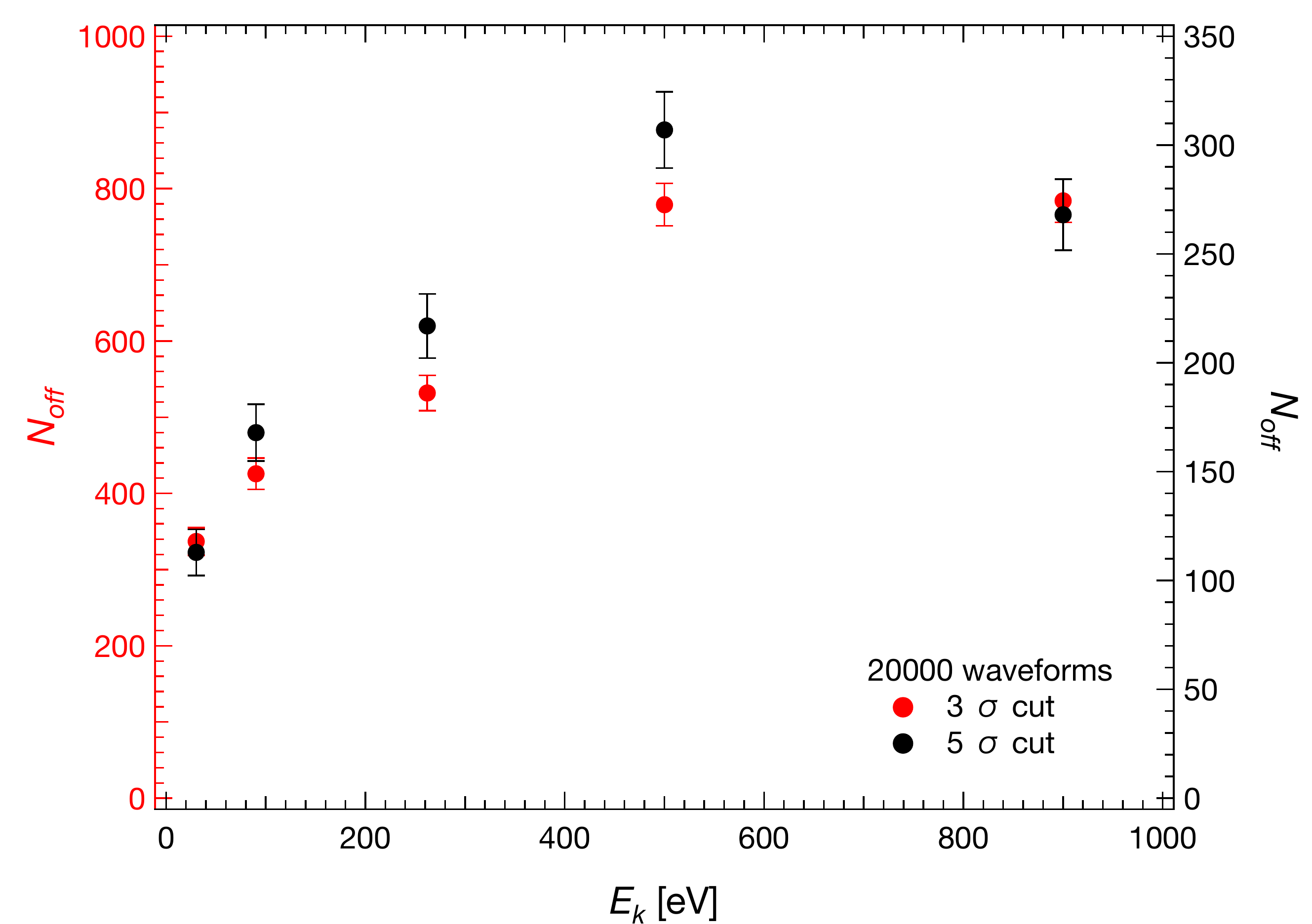}
\caption{Number of off-diagonal points $N_{off}$ as a function of the electron energy $E_k$. The two colours of the markers and the scales refer to a $3 \sigma$ cut-off (red) and a $5 \sigma$ cut-off (black).}
\label{fig:doppiVSenergia}
\end{figure}

The pulse height distribution has been characterised also at increasing beam current, by taking samples of 1000 waveforms and electrons within 30-900 eV. At all energies $E_k$, the mean amplitude of the PHD decreases when the beam current $I_{in}$ grows (see Supplemental material). The reduction of the mean amplitude corresponds to a \emph{shift} of the pulse height distribution toward the threshold level, therefore a growing number of events is cut off by the discriminator and the output count rate decreases. This behaviour leads to a saturation effect in the absolute detection efficiency of the MCP, as discussed in Section \ref{sec:efficiency}. \\
A further analysis of the pulse shape revealed a correlation between amplitude, area and width. We measured with the oscilloscope these three pulse characteristics on a sample of 20000 waveforms in a time window of $\tau$ = 10 ns. The beam current was approximately 10 fA, which corresponds to an output count rate $N_{out} \approx 30 \cdot 10^3\ s^{-1}$, and the measurements have been repeated for electrons with $E_k$ = 32, 92, 264, 502, 902 eV. In Figure \ref{fig:Correlations}, the peak area is plotted as a function of the amplitude and the colour scale represents the peak width. The majority of the points are aligned along a straight line (we call it \emph{diagonal}), while a few $\%$ are off-diagonal. The curves have been fitted with linear functions, shown as solid lines. Furthermore, the aligned points have an almost constant width of approximately 3 ns and the off-diagonal ones are either wider peaks or, a few of them, two time-separated pulses within the same window (red open markers).\\
We found this behaviour at all energies within the investigated range and also increasing output counting rate. The angular coefficient of the diagonal as a function of the beam energy, in turn, has been fitted with a straight line (see Supplemental material). The fit returned an intercept equal to 3.49 ns and an angular coefficient equal to $2.72 \cdot 10^{-14}$ s/eV. From the last result, it is possible to infer that the slope of the diagonal has no evident energy dependence. To double check this result, we fitted the angular coefficients curve also with a constant function and we found a constant value equal to 3.50 ns, very close to the linear fit result. In conclusion, the pulse shape of the MCP does not depend on the energy of the incoming electrons and their rate.\\
In order to understand the nature of the off-diagonal points shown in Figure \ref{fig:Correlations}, we studied the dependence on energy of these events at fixed output rate $N_{out}\sim 35 \cdot 10^3\ s^{-1}$, by establishing a method to count the off-diagonal points. For each point of the distribution reported in Figure \ref{fig:Correlations} we subtracted to the measured area the fitted intercept of the diagonal and then we divided by the measured amplitude. The resulting values are mainly distributed around a constant but the ones corresponding to off-diagonal points. Finally, by setting a cut-off on this distribution, we counted the points beyond it. In Figure \ref{fig:doppiVSenergia}, the number of off-diagonal points $N_{off}$ is shown as a function of the electrons energy $E_k$ for a $3 \sigma$ cut-off (red) and a $5 \sigma$ cut-off (black), where $\sigma$ is the standard deviation of the distribution. Although the choice of the cut-off is arbitrary, the energy dependence for the two cases is compatible within the experimental uncertainty: the number of off-diagonal points reaches a maximum at 502 eV and then decreases, or at least has a plateau, at 902 eV. This energy behaviour suggest that the off-diagonal events can be identified as double electrons which enter the MCP very close in time producing a single wide peak, or sufficiently separated to result in two resolved pulses. The incoming electrons can in fact strike the inter-channel surface and produce secondaries, which eventually enter adjacent channels \cite{Schagen, Wiza, Fraser}. This process is favoured by the presence of an electric field at the MCP surface \cite{Fraser}. This is not the case in our set-up, as the MCP front and the electron gun output are both grounded. As a consequence, the number of these events is small with respect to the total output rate, at 502 eV with $3 \sigma$ cut-off it is less than 3$\%$. On the contrary, when an electric field is applied the double events become an important fraction of the total events ($\sim 20\%$ \cite{Fraser}). The analysis of the output waveform described here provides a simple method to discriminate single- from multi- electron events.

\section{Absolute efficiency} 
\label{sec:efficiency}
\begin{figure}[h]
\centering
\includegraphics[scale=0.3]{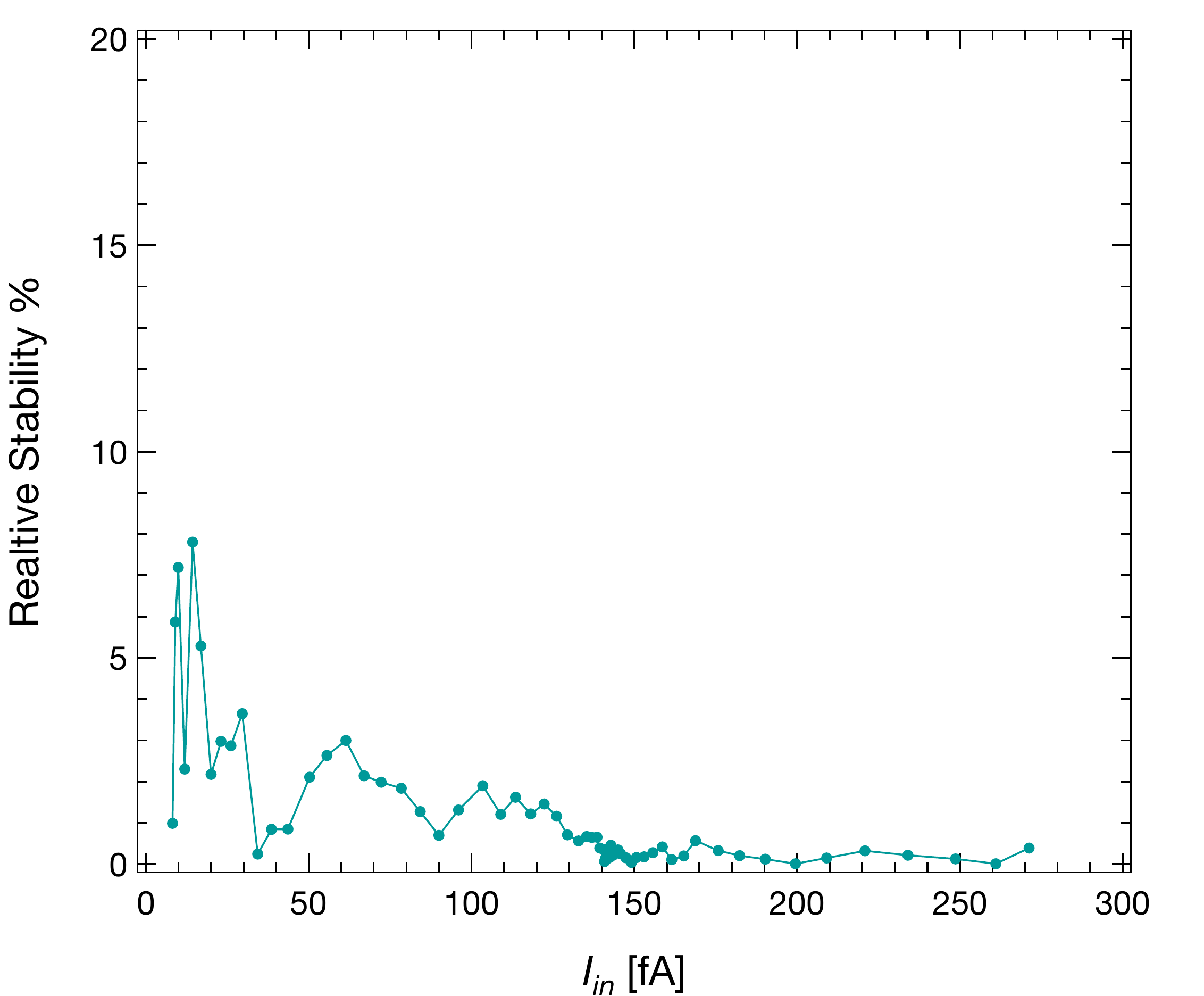}
\caption{Relative stability of the beam current (defined in the text).}
\label{fig:stability}
\end{figure}

To quantify the absolute efficiency of the MCP we sent an increasing electron current $I_{in}$ on the MCP surface and we read the count rate $N_{out}$ at the output of the device. The electron beam current was gradually varied approximately from $I_{in}$ = 5 fA to $I_{in}$ = 280 fA by adjusting the polarisation of one electrode in the source zone of the electron gun (we call this procedure \emph{current scan}).\\ 
The measurement comprised a first current scan on the Faraday cup, then the same current scan on the MCP, reading the output count rate, and again on the Faraday cup. In this procedure we used two beam sizes: $\sim$1 mm diameter when the electrons were sent on the Faraday cup and $\sim$4 mm diameter for the current scan on the MCP. The hole of the Faraday cup is in fact smaller (diameter 3 mm) than the MCP effective surface (diameter 14 mm). The choice of two sizes is necessary in order to have a beam which fits into the Faraday cup hole and a lower current density on the MCP, as an higher efficiency is expected and damages are avoided.\\
The repeated measurement (time interval approximately 20 minutes) of the current before and after the scan on the MCP is crucial, as it allows to check the stability of the beam current and to evaluate the uncertainty on the current measurement. In Figure \ref{fig:stability}, the relative stability of the current, defined as the difference of the two current measurements on the Faraday cup over the first measurement, is shown for a beam of energy 92 eV at increasing beam current $I_{in}$. In the shown case, the stability is approximately 1.2$\%$ on average and, above 130 fA, is less than 1$\%$. These are the typical values of the stability for all beam conditions investigated in this paper. The uncertainty on the current measurement is then given by the squared sum of the stability and the picoammeter accuracy, which is 1$\%$ + 3 fA in this measurement range. Thus the majority of the uncertainty is due to the accuracy of the instrument.

\begin{figure}[h]
\centering
\includegraphics[scale=0.29]{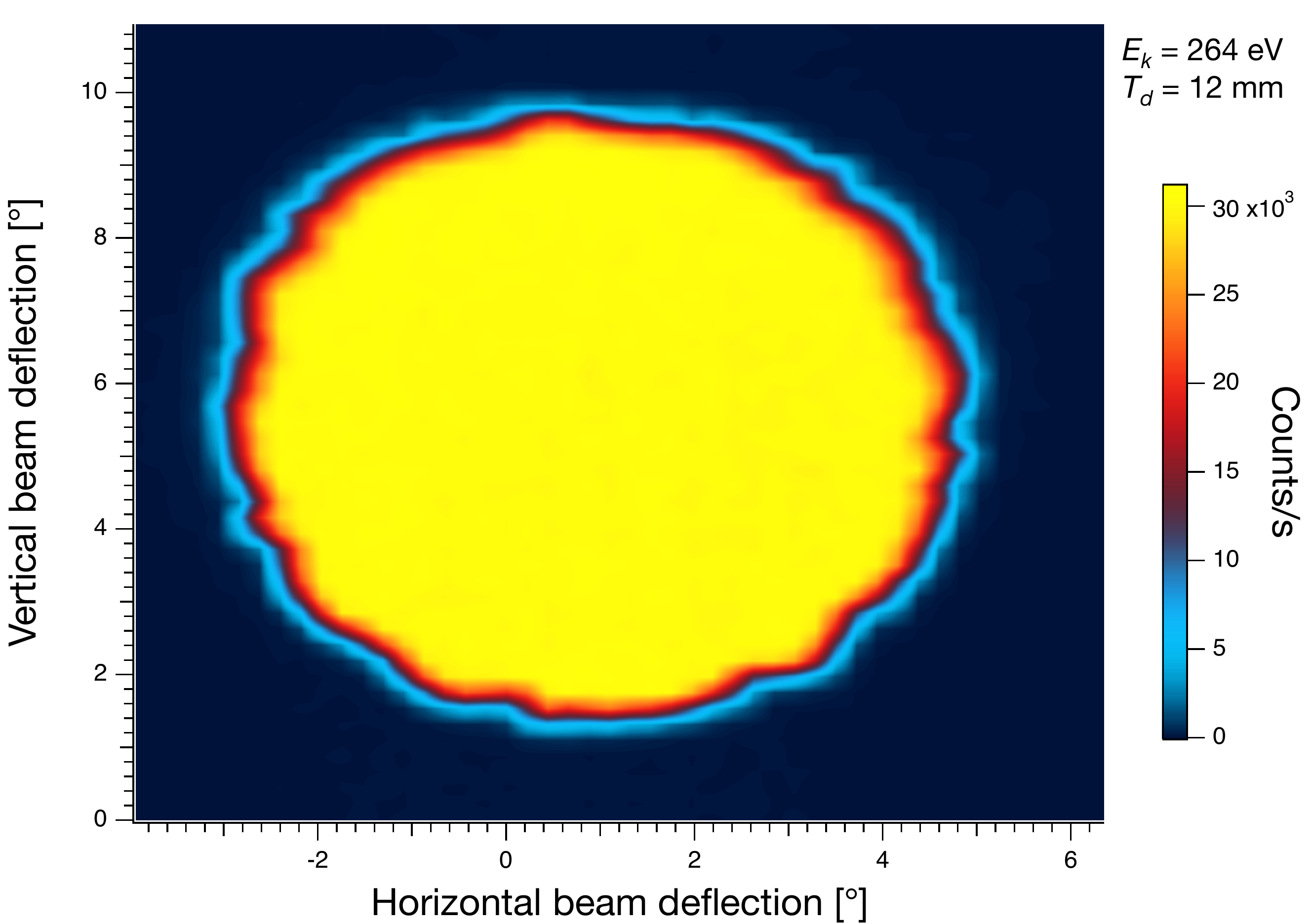}
\caption{MCP count rate as a function of the deviation angle of the beam with respect to the electron gun axis in two orthogonal directions.}
\label{fig:MCPangle}
\end{figure}

The incidence angle of the electron beam on the MCP effective surface can be critical for the efficiency, because an electron entering parallel to the axis channel, or with a small angle, can hit the walls far from the entrance resulting in a lower final amplification. However, in our specific set-up, we found a constant efficiency of the MCP as a function of the incidence angle. We performed a 2D map of the MCP surface with the electron beam at energy 264 eV and current approximately 7 fA. The map, shown in Figure \ref{fig:MCPangle}, represents the MCP count rate as a function of the beam deflection in two orthogonal directions. The deflection can be translated either in terms of spatial displacement of the beam spot or, as in this case, in terms of deviation angle of the beam with respect to the electron gun axis. From geometrical considerations, the incidence angle of the beam with respect to the MCP surface normal can be obtained. 

\begin{figure}[h]
\centering
\includegraphics[scale=0.35]{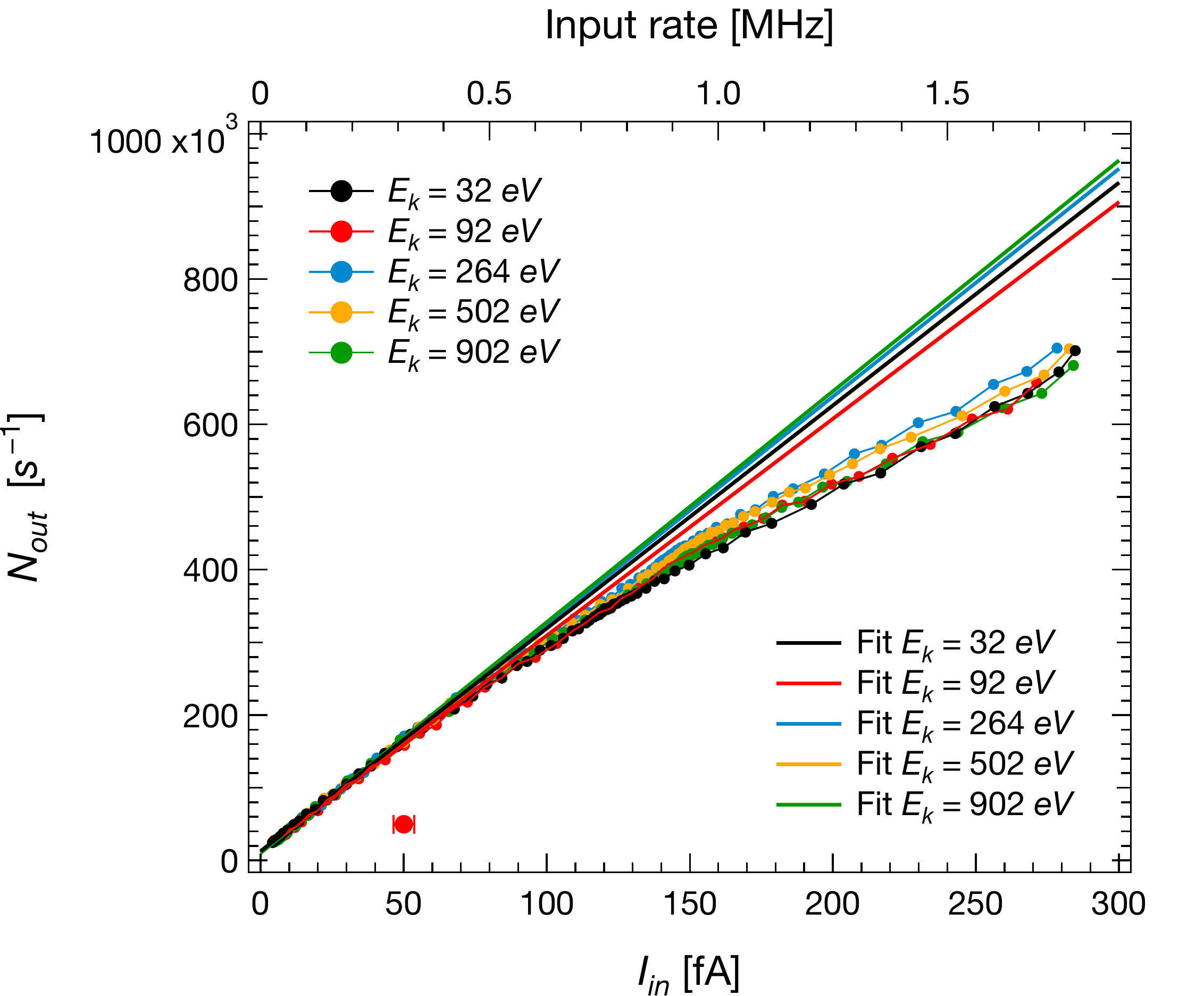}
\caption{Output count rate of the MCP as a function of the input beam current (lines with markers) and the linear fit (solid lines). The different colours are for different energies $E_k$ of the electron beam. The single red dot is an indication of the uncertainty associated to the measurements, not shown on all markers for sake of clarity.}
\label{fig:efficiency}
\end{figure}

We define the absolute efficiency $\epsilon$ of the MCP as

\begin{equation}
\label{eq:efficiency}
\epsilon = N_{out} \cdot  \frac{e}{I_{in}} 
\end{equation}

where $N_{out}$ is the rate of electrons at the MCP output, $I_{in}$ the incoming current of electrons and $e$ the electron charge. \\
In Figure \ref{fig:efficiency}, the measurements of the MCP output count rate as a function of the input beam current are shown as lines with markers. 

\begin{figure}[h]
\centering
\includegraphics[scale=0.33]{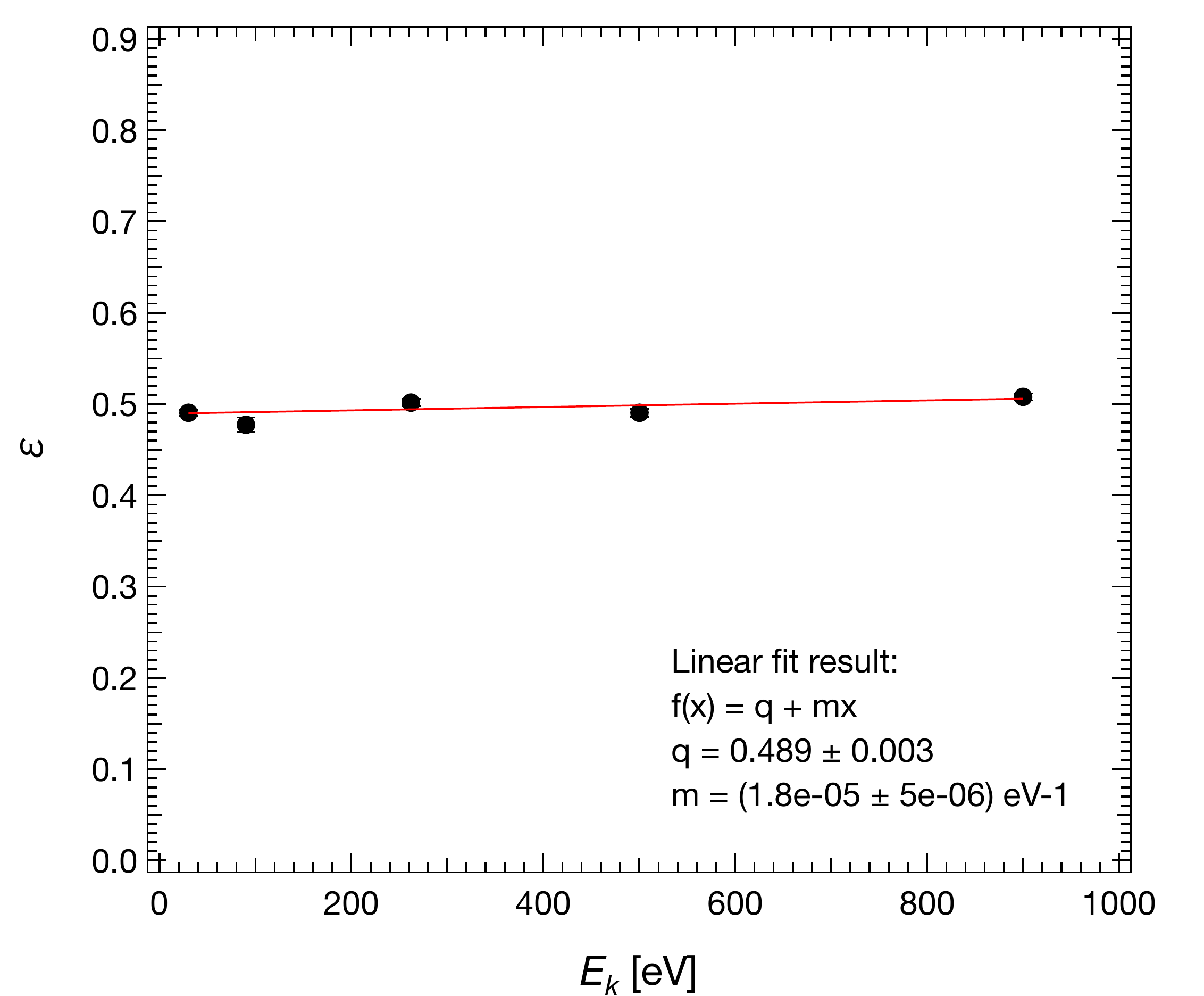}
\caption{Absolute efficiency as a function of the electron energy.}
\label{fig:efficiencyVSenergy}
\end{figure}

The beam current range used for the efficiency characterisation was from a few fA to approximately 280 fA, which corresponded, in terms of electron rate, to a few $10^4$ Hz up to less than 2 MHz. The different colours refer to the energies $E_k$ of the electron beam set for the measurement, in the 30-900 eV range. A single red dot is shown as an indication of the uncertainty associated to the measurements. At all energies, the trend of the output count rate has a linear behaviour in the low current region and a saturation above approximately 70 fA. As the input current increases, the gain level is high and the wall charge is not completely restored before the next electron enters the channel \cite{Schagen}. This mechanism leads to a decrease of the mean pulse amplitude, then a growing number of pulses are lost below the discriminator threshold and the output counting rate is reduced.\\
From a linear fit (solid lines) up to 50 fA on each curve, the absolute efficiency at the different energies in the experimental range has been obtained as the angular coefficient multiplied by the electron charge (\ref{eq:efficiency}). The obtained MCP absolute efficiency $\epsilon$ is shown in Figure \ref{fig:efficiencyVSenergy} as a function of the electron energy $E_k$. With a linear fit $f(x) = q + mx$ (red line) we found that the efficiency is $\epsilon = q = (0.489 \pm 0.003)$ and it shows no evident energy dependence, as the angular coefficient is $m = (1.8\cdot10^{-5} \pm 0.5\cdot10^{-5})$ eV$^{-1}$. \\

The upper limit of the detection efficiency is geometrically set by the open area ratio, which is $\sim$60$\%$ for the MCP we tested. However, this limit can be exceeded if secondaries are produced on the inter-channel surface and they enter adjacent channels \cite{Schagen, Wiza, Fraser}. This mechanism depends on the primary electron energy and actually seems to dominate the energy dependence of the absolute efficiency, as Fraser pointed out \cite{Fraser} applying his theoretical model to the experimental data reported by Galanti \emph{et al.} \cite{Galanti}. In their experimental set-up the income of secondaries is favoured by the presence of an electric field. Since no electric field is present at the MCP input, the absence of a clear energy dependence in our results is a consequence of the low fraction of detected secondaries, below 3$\%$ as shown in Section \ref{sec:pulseshape}. Although the number of these secondaries is not sufficient to contribute to the detection efficiency, the energy dependence of the off-diagonal events (shown in Section \ref{sec:pulseshape}) is compatible with the theoretical energy behaviour calculated by Fraser \cite{Fraser} for the contribution of the secondaries to the efficiency.

\section{Conclusions}
With our experimental apparatus it is possible to measure the absolute efficiency of an electron detector within the 30 - 900 eV energy range. In this paper, the characterisation of a two-stage chevron MCP is reported. With approximately 10 fA of input electron current, the pulse height distribution of the signal has a quasi-Gaussian shape, with a well visible peak above the noise level. A weak energy resolution has been found within the experimental range, with a 68$\%$ variation of the average pulse height between 32 and 502 eV. With a further analysis of the pulse shape, a method to discriminate single- and double- electron events in a time window of 10 ns has been found. The relation between the MCP output count rate and the input electron beam current presents a linear behaviour from 5 fA to approximately 70 fA and a saturation at higher currents. We found an absolute efficiency of $0.489 \pm 0.003$ with no evident energy dependence. This behaviour is due to the fact that, in absence of an attracting electric field, the secondary electrons produced on the MCP surface rarely enter the channels of the device.\\
The results reported in this work are planned to be employed in the context of new experiments as the development of an advanced UV light detector, a novel concept of light dark matter detector \cite{Cavoto, Apponi:2021lyd, Cavoto:2019flp}, and a spectrometer for the tritium $\beta$ spectra investigation \cite{PTOLEMY3, PTOLEMY4}.

\ack
We are grateful to Franco Marinilli for his technical support. The financial support of the ATTRACT project funded by the EC under Grant Agreement 777222 is acknowledged. We acknowledge INFN CNS2 support.

\section*{References}
\bibliographystyle{JHEP-mod}
\bibliography{bibliographyMCP}

\providecommand{\href}[2]{#2}\begingroup\raggedright\begin{thebibliography}{10}

\bibitem{PTOLEMY3}
M.~G. Betti, M.~Biasotti, A.~Bosc{\'{a}}, F.~Calle, J.~Carabe-Lopez, G.~Cavoto
  et~al., \emph{A design for an electromagnetic filter for precision energy
  measurements at the tritium endpoint},
  \href{https://doi.org/https://doi.org/10.1016/j.ppnp.2019.02.004}{\emph{Progress
  in Particle and Nuclear Physics} {\bfseries 106} (2019) 120}.

\bibitem{PTOLEMY4}
M.~G. Betti, M.~Biasotti, A.~Bosc{\'{a}}, F.~Calle, N.~Canci, G.~Cavoto et~al.,
  \emph{Neutrino physics with the {PTOLEMY} project: active neutrino properties
  and the light sterile case},
  \href{https://doi.org/10.1088/1475-7516/2019/07/047}{\emph{Journal of
  Cosmology and Astroparticle Physics} {\bfseries 2019} (2019) 047}.

\bibitem{Cavoto}
G.~Cavoto, F.~Luchetta and A.~Polosa, \emph{Sub-GeV dark matter detection with
  electron recoils in carbon nanotubes},
  \href{https://doi.org/10.1016/j.physletb.2017.11.064}{\emph{Physics Letters
  B} {\bfseries 776} (2018) 338}.

\bibitem{Apponi:2021lyd}
A.~Apponi, G.~Cavoto, C.~Mariani, F.~Pandolfi, I.~Rago, A.~Ruocco et~al.,
  \emph{Carbon nanostructures for directional light dark matter detection},
  {\emph{Proceedings of Science} {\bfseries 390} (2021) }.

\bibitem{Cavoto:2019flp}
G.~Cavoto, M.~G. Betti, C.~Mariani, F.~Pandolfi, A.~D. Polosa, I.~Rago et~al.,
  \emph{{Carbon nanotubes as anisotropic target for dark matter}},
  \href{https://doi.org/10.1088/1742-6596/1468/1/012232}{\emph{J. Phys. Conf.
  Ser.} {\bfseries 1468} (2020) 012232}
  [\href{https://arxiv.org/abs/1911.01122}{{\ttfamily 1911.01122}}].

\bibitem{Cocco}
A.~G. Cocco, G.~Mangano and M.~Messina, \emph{Probing low energy neutrino
  backgrounds with neutrino capture on beta decaying nuclei},
  \href{https://doi.org/10.1088/1742-6596/110/8/082014}{\emph{Journal of
  Physics: Conference Series} {\bfseries 110} (2008) 082014}.

\bibitem{APD}
A.~Apponi, G.~Cavoto, M.~Iannone, C.~Mariani, F.~Pandolfi, D.~Paoloni et~al.,
  \emph{Response of windowless silicon avalanche photo-diodes to electrons in
  the 90{\textendash}900 {eV} range},
  \href{https://doi.org/10.1088/1748-0221/15/11/p11015}{\emph{Journal of
  Instrumentation} {\bfseries 15} (2020) P11015}.

\bibitem{Tulej}
M.~Tulej, S.~Meyer, M.~L\"uthi, D.~Lasi, A.~Galli, L.~Desorgher et~al.,
  \emph{Detection efficiency of microchannel plates for $e^-$ and $\pi^-$ in
  the momentum range from 17.5 to 345 MeV/c},
  \href{https://doi.org/10.1063/1.4928063}{\emph{Review of Scientific
  Instruments} {\bfseries 86} (2015) 083310}.

\bibitem{Kennerly}
R.~E. Kennerly, \emph{High-resolution pulsed electron beam time-of-flight
  spectrometer}, \href{https://doi.org/10.1063/1.1134931}{\emph{Review of
  Scientific Instruments} {\bfseries 48} (1977) 1682}.

\bibitem{Dhawan}
S.~Dhawan and R.~Majka, \emph{Development Status of Microchannel Plate
  Photomultipliers}, \href{https://doi.org/10.1109/TNS.1977.4328688}{\emph{IEEE
  Transactions on Nuclear Science} {\bfseries 24} (1977) 270}.

\bibitem{FraserXray}
G.~W. Fraser and J.~F. Pearson, \emph{Soft X-ray energy resolution with
  microchannel plate detectors of high quantum detection efficiency},
  \href{https://doi.org/10.1016/0167-5087(84)90156-X}{\emph{Nuclear Instruments
  and Methods in Physics Research} {\bfseries 219} (1984) 199}.

\bibitem{Galanti}
M.~Galanti, R.~Gott and J.~F. Renaud, \emph{A High Resolution, High Sensitivity
  Channel Plate Image Intensifier for Use in Particle Spectrographs},
  \href{https://doi.org/10.1063/1.1685013}{\emph{Review of Scientific
  Instruments} {\bfseries 42} (1971) 1818}.

\bibitem{Muller}
A.~M\"uller, N.~Djuri\'c, G.~H. Dunn and D.~S. Beli\'c, \emph{Absolute
  detection efficiencies of microchannel plates for 0.1-2.3 keV electrons and
  2.1-4.4 keV $Mg^+$ ions},
  \href{https://doi.org/10.1063/1.1138944}{\emph{Review of Scientific
  Instruments} {\bfseries 57} (1986) 349}.

\bibitem{Goruganthu}
R.~R. Goruganthu and W.~G. Wilson, \emph{Relative electron detection efficiency
  of microchannel plates from 0-3 keV},
  \href{https://doi.org/10.1063/1.1137709}{\emph{Review of Scientific
  Instruments} {\bfseries 55} (1984) 2030}
  [\href{https://arxiv.org/abs/https://doi.org/10.1063/1.1137709}{{\ttfamily
  https://doi.org/10.1063/1.1137709}}].

\bibitem{DiFilippo_Liscio_Ruocco_2020}
G.~{Di Filippo}, A.~Liscio and A.~Ruocco, \emph{The evolution of hydrogen
  induced defects and the restoration of $\pi$-plasmon as a monitor of the
  thermal reduction of graphene oxide},
  \href{https://doi.org/https://doi.org/10.1016/j.apsusc.2020.145605}{\emph{Applied
  Surface Science} {\bfseries 512} (2020) 145605}.

\bibitem{Paoloni_DiFilippo_Cvetko_Kladnik_Morgante_Ruocco_2020}
D.~Paoloni, G.~Di~Filippo, D.~Cvetko, G.~Kladnik, A.~Morgante and A.~Ruocco,
  \emph{Strong Chemical Interaction and Self-Demetalation of
  Zinc-Phthalocyanine on Al(100)},
  \href{https://doi.org/10.1021/acs.jpcc.0c06980}{\emph{The Journal of Physical
  Chemistry C} {\bfseries 124} (2020) 22550}
  [\href{https://arxiv.org/abs/https://doi.org/10.1021/acs.jpcc.0c06980}{{\ttfamily
  https://doi.org/10.1021/acs.jpcc.0c06980}}].

\bibitem{Schagen}
P.~Schagen, \emph{{Image tubes with channel electron multiplication}},  in
  \emph{{Advances in image pickup and display. Vol.\_1.}}, B.~{Kazan}, ed.,
  Academic Press, New York, (1974).

\bibitem{Wiza}
J.~Ladislas~Wiza, \emph{Microchannel plate detectors},
  \href{https://doi.org/10.1016/0029-554X(79)90734-1}{\emph{Nuclear Instruments
  and Methods} {\bfseries 162} (1979) 587}.

\bibitem{Fraser}
G.~Fraser, \emph{The electron detection efficiency of microchannel plates},
  \href{https://doi.org/https://doi.org/10.1016/0167-5087(83)90381-2}{\emph{Nuclear
  Instruments and Methods in Physics Research} {\bfseries 206} (1983) 445}.

\end{thebibliography}\endgroup

\end{document}